\documentclass{Interspeech2024}

\usepackage[table,xcdraw]{xcolor}
\usepackage{tikz,cite}
\usetikzlibrary{automata,positioning,shapes}
\interspeechcameraready 

\usepackage{subcaption}
\usepackage[bottom]{background}
\SetBgContents{Proceedings of Interspeech 2024. 1-5 September 2024, Kos, Greece. doi: 10.21437/Interspeech.2024-74}
\SetBgVshift{3.0cm}
\SetBgHshift{0.0cm}
\SetBgColor{black}
\SetBgOpacity{1.0}
\SetBgAngle{0.0}
\SetBgScale{1.0}

\newcommand{\alignnet}{AlignNet}
\newcommand{\audionet}{AudioNet} 
\newcommand{\alignmentnet}{Aligner} 

\title{AlignNet: Learning dataset score alignment functions to enable better training of speech quality estimators}
\name[affiliation={1}]{Jaden}{Pieper}
\name[affiliation={1}]{Stephen}{Voran}

\address{
  $^1$Institute for Telecommunication Sciences, National Telecommunications and Information Administration, United States}
\email{jpieper,svoran@ntia.gov}

\keywords{corpus effect, listening experiment, machine learning, no-reference estimator, speech quality, subjective test}

\begin{document}

\maketitle

\begin{abstract}
    
We develop two complementary advances for training no-reference (NR) speech quality estimators with independent datasets.
Multi-dataset finetuning (MDF) pretrains an NR estimator on a single dataset and then finetunes it on multiple datasets at once, including the dataset used for pretraining.
\alignnet{} uses an \audionet{} to generate intermediate score estimates before using the \alignmentnet{} to map intermediate estimates to the appropriate score range.
\alignnet{} is agnostic to the choice of \audionet{} so any successful NR speech quality estimator can benefit from its \alignmentnet{}.
The methods can be used in tandem, and we use two studies to show that they improve on current solutions: one study uses nine smaller datasets and the other uses four larger datasets.
\alignnet{} with MDF improves on other solutions because it efficiently and effectively removes misalignments that impair the learning process, and thus enables successful training with larger amounts of more diverse data.
\end{abstract}

\vspace{-2mm}
\section{Background}
\label{sec:background}
\vspace{-1mm}
Speech quality, naturalness, and related quantities can be measured in listening experiments or estimated by algorithms.
Algorithms can save much time and effort, but it is a significant challenge to develop an algorithm that produces reliable estimates across a wide range of conditions~\cite{Cooper2021}.
The most useful algorithms are called ``no-reference'' (NR) because they measure impaired speech directly and do not need a reference speech signal for comparison.
NR algorithms parallel absolute category rating (ACR) listening experiments where listeners score impaired speech without comparing to reference speech.

NR estimators originally used explicit models
\cite{RFK1994,ANIQUE,P.563,falkSRMRforNR} but moved to data-driven implicit modeling as machine learning (ML) became more mature and practical. A few examples include \cite{Soni2016,Hakami2017,Fu2018,Gamper2019,ReddyICASSP2022,Ryandhimas2023,wawenetsAccess}.
The ML approach is powerful and effective, but also highly dependent on the quantity and diversity of listening experiment results that comprise the ground-truth training data.
This motivates us to combine the results of multiple listening experiments to achieve the needed quantity and diversity.

\vspace{-2mm}
\subsection{The alignment problem}
\label{ssec:alignment}
\vspace{-1mm}
Despite the name, ACR results are not truly absolute.  They can depend on a variety of factors including characteristics of individual listeners and the range of conditions included in an experiment \cite{P8002,P1401,cooperIS23,lemaguer22_interspeech}. This is sometimes called the ``corpus effect.''
For example, in \cite{cooperIS23} a single fixed synthesized voice file received quality scores ranging from 1.8 to 4.5 in five different experiments, where the only difference between those experiments was the range of conditions included in each. 
More generally, for one of the five experiments in \cite{cooperIS23} 41\% of the speech files move by more than 1.0 on the five-point MOS scale when rated in one of the other four experiments.  
This behavior can be attributed to the listeners' desire to use the entire scale.

The corpus effect creates a problem when we seek to combine results of multiple experiments.
Naive combining can yield inconsistent training data that harms training processes instead of enhancing them.
But if the results of each experiment can be brought to a single common scale (``aligned''), they can then work together to improve training of an NR estimator.
This requires finding a useful common scale and an optimal mapping from each set of experiment results to that common scale.
This is the dataset alignment problem.

\vspace{-2mm}
\subsection{Prior work and our contributions}
\label{ssec:prior}
\vspace{-1mm}
Several alignment techniques have been proposed and used.
When two listening experiments include common conditions, results for these conditions may be used to develop alignment functions.
Historically many experiments included a standardized adjustable reference condition called the modulated noise reference unit (MNRU) \cite{p810} for exactly this purpose.
This is currently not common practice, likely because the impairments produced by the MNRU sound very different from impairments appearing in current experiments, thus limiting its usefulness as a reference condition.  Other standardized reference conditions have been used in the past but devoting experiment conditions to references always consumes precious resources.

An iterative approach that alternately optimizes alignment functions and an estimation algorithm is given in \cite{INLSA}.  An updated iterative approach that leverages ML is given in \cite{MittagQomex2021}. Other approaches explore individual alignments for each listener in a listening test
\cite{Nessler2021IS,Leng2021,Huang22}.
Aligning listeners can compensate for their individual behaviors and can lead to better training of NR estimators. 
But this cannot account for the primary portion of the corpus effect, caused by the broader biases due to each experiment's context. 
Further, listener alignment is not possible unless datasets label each listener's scores.

In this paper we offer the following novel contributions:
\begin{itemize}
    \item multi-dataset finetuning, a progressive training regimen that advantageously leverages both larger and smaller datasets
    \item adding a small score alignment network and a dataset indicator to an audio network
    \item combining these to learn embeddings for the dataset indicator, alignment functions, and optimal audio network weights
    \item using an unprecedented 13 datasets covering 3 languages, scores for 4 different speech attributes,  and a very wide range of measurement domains, totalling over 300 hours of speech
    \item demonstrating that these innovations allow previously incompatible datasets to collaborate during training, resulting in better estimates across disparate measurement domains.
\end{itemize}

\vspace{-2mm}
\section{No-reference speech quality estimators for multiple datasets}
\label{sec:NR}
\vspace{-1mm}
Here we discuss issues with the conventional strategy for training a speech quality estimator using multiple datasets. 
We then propose two innovations that enable learning meaningful relationships between audio and scores across multiple datasets, even when inconsistent scores are present.

\vspace{-2mm}
\subsection{Conventional approach}\label{conventional}
\vspace{-1mm}
The conventional approach to training an NR speech quality estimator with speech and scores from multiple distinct listening experiments is to simply use all the datasets at once.
However, due to the corpus effect, there can be a misalignment between speech and target scores from different experiments.
When identical or very similar speech files appear in multiple listening experiments, they almost certainly receive different scores in each experiment.
This means that while training, the network must attempt to map identical or similar  input files to multiple conflicting output scores, and would likely estimate a score that is roughly the average of all scores seen for the file.
This is reasonable to an extent, but these disparate scores for the same input add additional noise for the network to sift through and place inherent limitations on its estimates;
it can never produce a single estimate for this input that achieves low loss for all the associated target scores.

\vspace{-2mm}
\subsection{Multi-dataset finetuning}\label{pretrain-finetune}
\vspace{-1mm}
In the conventional approach the network attempts to learn audio relationships while dealing with misaligned target scores from multiple experiments, which impedes the training process.
Previous work has demonstrated the benefits of pretraining a network on one dataset and then switching to a different dataset for finetuning~\cite{Cooper2021,Tseng2021,Becerra2022}.
The initial pretraining allows the network to learn a basic and somewhat transferable relationship between audio and scores.
Here we propose multi-dataset finetuning (MDF), where we first pretrain the network on a single dataset before finetuning with all the datasets at once, including the original dataset used for pretraining.
Pretraining places the network into a state where it already knows some meaningful relationships between audio and scores, and can then  balance the misaligned scores from the different listening experiments.
It has some of the same limitations as the conventional approach, but pretraining on a single dataset enables much better training and predictions when applied to multiple datasets.
We believe MDF is a novel approach to the problem.

\vspace{-2mm}
\subsection{Dataset alignment with \alignnet}\label{alignnet}
\begin{figure}[h]
	\begin{center}
		\begin{tikzpicture}[->, auto, thick, node distance=1.2cm]
    \tikzstyle{every state}=[fill=white,draw=black,thick,text=black,scale=1,rectangle]
    \node[state] (Audio)[ellipse] {Audio};
    \node[state] (Dataset)[ellipse,right=0.2cm of Audio] {\begin{tabular}{c} Dataset \\ indicator\end{tabular}};
    \node[state] (Feat)[dashed,above of=Audio] {Feature extractor};
    \node[state] (AudioNet)[above of=Feat] {\audionet};
    \node[state] (Alignment)[above=of AudioNet]{\alignmentnet};
    \node[state] (MOS)[ellipse,right=0.5cm of Alignment]{\begin{tabular}{c} Score estimates \end{tabular}};
    \path
    (Audio) edge (Feat)
    (Feat) edge (AudioNet)
    (AudioNet) edge (Alignment.south)
    (Alignment) edge (MOS)
    ;
    \draw
    (Dataset) ->(2.2,3.75) -| ([xshift=3mm] Alignment.south)
    ;
\end{tikzpicture}
		\caption{AlignNet model diagram. The feature extractor is optional depending on the choice of \audionet{}.}
		\label{fig:AlignNet}
	\end{center}
\vspace{-8mm}
\end{figure}
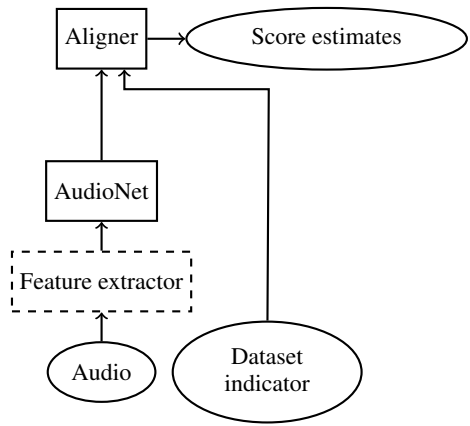

\vspace{-1mm}
We now introduce a novel architecture called \alignnet{}, which allows any NR speech quality estimator to better benefit from multiple datasets, with only a minimal increase in network complexity.
\alignnet{} is essentially two network components in sequence, which we call the \audionet{} and the \alignmentnet{} respectively.
\alignnet{} is intentionally designed to be agnostic to the choice of the \audionet, which maps audio or audio features (depending on the choice of \audionet{}) to intermediate score estimates. The \alignmentnet{} uses a categorical dataset indicator to map those intermediate score estimates to final scores for the appropriate dataset.
The network architecture is outlined in Fig.~\ref{fig:AlignNet}.

It is necessary to select a reference dataset and force the \alignmentnet{} to apply the identity function to  reference dataset results.  This ensures that the outputs of \audionet{} are grounded in meaningful quality scores, and that \audionet{} gives estimates for all audio in the domain of the reference dataset scores.
Favorable attributes for a reference dataset include trustworthy scores and a wide range of conditions. This allows \audionet{} to produce outputs on a level playing field for comparisons of speech files, without any corpus effect biases muddling the rankings.
Finally this ensures that the mappings learned by the \alignmentnet{} carry relevant information about the relationships between different experiments and are easily interpretable.

To build an intuition, again consider multiple listening experiments where very different scores were reported for an identical condition and audio file. 
In \alignnet{}, the \audionet{} would give the same result for each occurrence of the audio file, and then the \alignmentnet{} could use the dataset indicator to successfully map this single value to the appropriate, different scores. Thus \alignnet{} is able to yield low loss for each occurrence of the input, in spite of their different scores.

The \alignmentnet{} is extremely light-weight and adds an insignificant number of parameters to any effective \audionet{}.
It first maps the dataset indicator to an $N$-dimensional embedding and concatenates that embedding with the \audionet{} estimation.
The network consists of a series of fully connected layers of identical dimensions, separated by ReLU activations, and a final fully connected layer that maps the data to a single score estimate.
The number of parameters in the \alignmentnet{} is dependent on the number of datasets used, and in our implementations it has roughly 1100 parameters total.
More in-depth implementation details are available with our source code\footnote{\url{https://github.com/NTIA/alignnet}}.

Successful training of \alignnet{} requires a clear division of labor --- the \audionet{} should not attempt to make alignments and the \alignmentnet{} should not attempt to measure audio.
This is achieved by MDF, the use of a reference dataset, and freezing the \audionet{} for one epoch at the start of finetuning.
The reference dataset is used for pretraining so \audionet{} is well-positioned at the beginning of training, at least for that dataset.
Continuing to train allows the \audionet{} to better learn the speech from the other datasets, while the \alignmentnet{} learns how to reconcile the scores across the different datasets.

\vspace{-2mm}
\subsection{Loss function}
\vspace{-1mm}
In conventional network training each individual piece of data is given equal weight in the loss function.
When training a network with multiple datasets, we propose that each \emph{dataset} be given equal weight within the loss function using
\vspace{-2mm}
\begin{equation}\label{eqn:weighted-loss}
    L = \frac{1}{N_d}\sum_{j=1}^{N_d} l(\mathbf{y}_j, \mathbf{\hat{y}}_j),
    \vspace{-2mm}
\end{equation}
where $N_d$ is the number of datasets used in training, $\mathbf{y}_j$ are all the targets for dataset $j$, $\mathbf{\hat{y}}_j$ are all the estimates for dataset $j$, and $l$ is mean-squared error loss.
This allows a network to achieve good results for each dataset, rather than letting larger datasets dominate the learning.

\vspace{-2mm}
\section{Experiments and Results}
\vspace{-1mm}
We performed studies with two different groups of datasets: one with nine smaller datasets that have more votes per file and one with four larger datasets that have fewer votes per file.
Key properties are summarized in Table \ref{table:datasets}. Three languages are represented and the total duration of speech exceeds 300 hours.
Diverse measurement domains include synthesized speech, voice conversion, neural vocoders, conventional codecs, packet loss, noise, reverb, enhancement, filtering, and more.
The NOIZEUS and PSTN datasets are narrowband (nominally 300 to 3600 Hz) and the remaining datasets support wideband (nominally 50 to 7000 Hz) or fullband speech. 
Scores are for four different attributes: ``Acceptability'' (B21 S1 Acc), ``Naturalness'' (remaining 4 Blizzard datasets, VMC22, and VCC18), ``Overall Quality'' (NOIZEUS), and ``Speech Quality'' (remaining 5 datasets).
These attributes are related but not identical which further motivates dataset alignment.
 
\begin{table}[tbp]
    \caption{Summary of 13 datasets used in this work. The first 9 have 8 to 24 votes per file, the final 4 have 4 or 5 votes per file on average.  Blizzard 2021 uses Spanish language, Tencent uses Chinese, all others use English.}
    \vspace{-1mm}
    \centering
    \resizebox{\columnwidth}{!}{
        \begin{tabular}{|l|l|l|r|}
\hline
\textbf{Dataset}                                                         & \textbf{Abbr.}                                            & \textbf{Domain}                                                                                                                  & \textbf{\begin{tabular}[c]{@{}l@{}}Number\\ of Files\end{tabular}}  \\ \hline
Blizzard   2021 SS1  \cite{Blizzard2021}                                             & B21   S1 Nat    & \begin{tabular}[c]{@{}l@{}}Synthesized \& \\natural speech \end{tabular}                                                 & 242                                                                                                                 \\ \hline
Blizzard   2021 SH1 \cite{Blizzard2021}                                              & B21   H1 Nat    & same as above                                               & 338                                                                                                                 \\ \hline
Blizzard   2021 SS1 \cite{Blizzard2021}                                              & B21   S1 Acc  & same  as above                                            & 363                                                                                                               \\ \hline
Blizzard   2008 News  \cite{Blizzard2008}                                            & B08   News     & same as above                                                & 802                                                                                                                \\ \hline
Blizzard   2008 Novels  \cite{Blizzard2008}                                          & B08   Novel   & same as above                                               & 802                                                                                                               \\ \hline
FFTnet   \cite{JinICASSP2018}                                                         & FFTnet            & Neural vocoders                                                                                                          & 1200                                                                                                               \\ \hline
NOIZEUS \cite{LoizouICASSP2006}                                                          & NOIZEUS                                                    & Noise \&  suppression                                                          & 1664                                                                                                                \\ \hline
\begin{tabular}[c]{@{}l@{}}VoiceMOS\\ Challenge 2022\cite{voicemos2022}\end{tabular} & VMC22                                                       & \begin{tabular}[c]{@{}l@{}}Speaker conversion  \& \\ synthesized speech\end{tabular}                                    & 7106                                                                                                                \\ \hline
Tencent  \cite{ConfSpeech2020}                                                         & Tencent                                                   & \begin{tabular}[c]{@{}l@{}}Noise, suppression,\\ reverb, coding,\\ packet loss \& concealment\end{tabular}   & 11,563                                                                                                             \\ \hline
\hline
NISQA SIM  \cite{Mittag2021IS}                                                         & NISQA                                                   & \begin{tabular}[c]{@{}l@{}}Coding, packet loss, \\ noise, filtering \& clipping\end{tabular}   & 12,500                                                                                                             \\ \hline

\begin{tabular}[c]{@{}l@{}}Voice Conversion \\ Challenge 2018 \cite{VCC2018} \end{tabular}                                                         & VCC18                                                 & Voice conversion systems   & 20,580                                                                                                              \\ \hline

Indiana U. MOS  \cite{Dong2020}                                                         & IU MOS                                                     & Noise \& reverb   & 36,000                                                                                                              \\ \hline

PSTN  \cite{Mittag2020Interspeech}                                                         & PSTN                                                    & \begin{tabular}[c]{@{}l@{}}PSTN to VoIP calls \\ plus noise\end{tabular}   & 58,709                                                                                                              \\ \hline
\end{tabular}
    }
\label{table:datasets}
\vspace{-4mm}
\end{table}
In each study we explored different training regimens and network architectures with multiple datasets in order to demonstrate the performance of our two novel approaches compared to existing methods.
We used the NR speech quality estimator MOSNet~\cite{Lo2019} for all experiments in both studies, either exclusively or as the \audionet{} in \alignnet{}.
We chose MOSNet as it is a sufficiently large network to have the capacity to achieve good results for these studies, while being small enough to train relatively quickly and be more usable in practice. Further it has been successfully used as a baseline model in related research on listener specific corrections~\cite{Leng2021,Huang22}.
Unlike the original MOSNet implementation, we did not opt to use frame level loss, and instead averaged frames into a single value before the loss function; otherwise our implementation exactly matches the original paper.
All audio was resampled to 16 kHz prior to the STFT calculation.
We randomly split each dataset 
into 80\%, 10\%, and 10\% for the
training, validation, and testing data respectively.
We use the same split across all tests to ensure a fair comparison, and all reported results are from the unseen test sets.
We trained all networks with the loss function defined in (\ref{eqn:weighted-loss}), except in the bias-aware loss comparison (BAL), which uses the loss function defined in~\cite{MittagQomex2021}.
In the small dataset study the Tencent dataset was selected as the reference dataset, which means we also used it for MDF pretraining.  In the large dataset study the NISQA dataset filled this role.

We use ``depth'' to describe network performance for a single dataset and ``breadth'' to describe performance across all datasets of interest.
Each \audionet{} has its natural tradeoff between depth and breadth; at a certain point one cannot be improved without harming the other.
Adding dataset alignment can mitigate this tradeoff and allow better simultaneous depth and breadth.
\alignnet{}'s \alignmentnet{} generally improves depths without reducing breadth.

We use two metrics to evaluate the depth and breadth performance in our experiments: Pearson's linear correlation coefficient (LCC) describes the networks ability to rank speech attributes, and root mean-squared error (RMSE) describes the distance of the network's estimates from the true scores.
The small and large dataset study results are given in Tables~\ref{table:small-corrs} and
~\ref{table:large-corrs}, respectively.
Each column gives results for the unseen testing portion of a given dataset.
Bold indicates the best and underlining the second best performance for each column.  
The * symbol denotes a statistically significant improvement over the conventional regimen (``All" row), calculated using Zou's confidence interval for LCC~\cite{zou2007} and bootstrapping for RMSE~\cite{efron1994}.
For the large dataset study full results are shown for training on each dataset and the diagonal is shaded; off-diagonal results show the lack of breadth.
We did the same for the small dataset study but for brevity the diagonal is compressed into the ``Individual" row, where each cell shows results for training and testing (on unseen data) for a given dataset.
No individually trained model shows any meaningful breadth
and none of the unshown values were first or second place for any column.
Note that it is more difficult to achieve statistical significance for the datasets with fewer than 500 files, as the test sets are very small.

\begin{table*}[tbp]
    \caption{
    LCC (above) and RMSE (below) for all models on the small datasets. 
        }
    \vspace{-1mm}
    \label{table:small-corrs}
    \centering
    \resizebox{\linewidth}{!}{
        \begin{tabular}{|c|c|c|c|c|c|c|c|c|c|c|} \hline
	\textbf{Training Data} & \textbf{B21 S1 Nat} & \textbf{B21 H1 Nat} & \textbf{B21 S1 Acc} & \textbf{B08 Novel} & \textbf{B08 News} & \textbf{FFTNet} & \textbf{NOIZEUS} & \textbf{VMC22} & \textbf{Tencent} & \textbf{All}\\ \hline
	Individual & 0.45 & 0.65 & 0.23 & 0.66 & 0.67 & \textbf{0.81*} & \textbf{0.80*} & 0.51 & \textbf{0.94*} & NA \\ \hline
\hline	All & 0.73 & 0.82 & \underline{0.70} & 0.62 & 0.69 & 0.53 & 0.65 & 0.71 & 0.80 & 0.77 \\ \hline
	All (+ BAL) & \underline{0.83} & 0.77 & 0.65 & 0.56 & 0.55 & 0.69* & 0.70 & \underline{0.75*} & 0.91* & 0.83* \\ \hline
	All (+ MDF) & 0.82 & \underline{0.83} & 0.62 & \underline{0.72} & \underline{0.77} & \underline{0.70*} & 0.71 & 0.74 & 0.89* & \underline{0.84*} \\ \hline
	All (+ MDF + AlignNet) & \textbf{0.88} & \textbf{0.90} & \textbf{0.78} & \textbf{0.81*} & \textbf{0.82*} & 0.66* & \underline{0.76*} & \textbf{0.76*} & \underline{0.92*} & \textbf{0.87*} \\ \hline
\end{tabular}
    }
    \begin{tabular}{cc}
         &  \\[-4mm]
    \end{tabular}
    \resizebox{\linewidth}{!}{
        \begin{tabular}{|c|c|c|c|c|c|c|c|c|c|c|} \hline
	\textbf{Training Data} & \textbf{B21 S1 Nat} & \textbf{B21 H1 Nat} & \textbf{B21 S1 Acc} & \textbf{B08 Novel} & \textbf{B08 News} & \textbf{FFTNet} & \textbf{NOIZEUS} & \textbf{VMC22} & \textbf{Tencent} & \textbf{All}\\ \hline
	Individual & 1.15 & 1.19 & 0.95 & 0.88 & 0.87 & 0.66* & \underline{0.34*} & 0.95 & \textbf{0.41*} & NA \\ \hline
\hline	All & 0.87 & \underline{0.69} & \underline{0.56} & 0.85 & 0.70 & 0.74 & 0.44 & 0.66 & 0.80 & 0.73 \\ \hline
	All (+ BAL) & \underline{0.65*} & 0.70 & 0.70 & 0.93 & 0.78 & \textbf{0.55*} & 0.40* & 0.66 & 0.58* & 0.62* \\ \hline
	All (+ MDF) & 0.79* & 0.74 & 0.68 & \underline{0.66*} & \underline{0.57*} & 0.62* & 0.35* & \underline{0.61*} & 0.55* & \underline{0.57*} \\ \hline
	All (+ MDF + AlignNet) & \textbf{0.54*} & \textbf{0.46*} & \textbf{0.56} & \textbf{0.55*} & \textbf{0.51*} & \underline{0.55*} & \textbf{0.33*} & \textbf{0.60*} & \underline{0.46*} & \textbf{0.51*} \\ \hline
\end{tabular}
    }
    \vspace{-3mm}
\end{table*}

\begin{table}[tbp]
    \caption{LCC (above) and RMSE (below) for all models on the large datasets. 
    }
    \vspace{-1mm}
    \label{table:large-corrs}
    \centering
    \resizebox{\columnwidth}{!}{
        \begin{tabular}{|c|c|c|c|c|c|} \hline
	\textbf{Training Data} & \textbf{NISQA} & \textbf{VCC18} & \textbf{IU} & \textbf{PSTN} & \textbf{All}\\ \hline
	NISQA & \cellcolor[HTML]{e6e6e6}0.89* & 0.48 & 0.69 & 0.74 & 0.39 \\ \hline
	VCC18 & 0.50 & \cellcolor[HTML]{e6e6e6} \textbf{0.66*} & 0.42 & 0.53 & 0.07 \\ \hline
	IU & 0.54 & 0.13 & \cellcolor[HTML]{e6e6e6} \textbf{0.97*} & 0.58 & 0.50 \\ \hline
	PSTN & 0.72 & 0.38 & 0.62 & \cellcolor[HTML]{e6e6e6} \textbf{0.81*} & 0.37 \\ \hline
\hline	All & 0.80 & 0.60 & 0.86 & 0.75 & 0.88 \\ \hline
	All (+ BAL) & 0.88* & 0.62 & 0.96* & 0.80* & \underline{0.94*} \\ \hline
	All (+ MDF) & \underline{0.91*} & 0.63* & 0.96* & \underline{0.81*} & 0.94* \\ \hline
	All (+ MDF + AlignNet) & \textbf{0.91*} & \underline{0.64*} & \underline{0.97*} & 0.80* & \textbf{0.94*} \\ \hline
\end{tabular}
    }
    \begin{tabular}{cc}
         &  \\[-4mm]
    \end{tabular}
    \resizebox{\columnwidth}{!}{
        \begin{tabular}{|c|c|c|c|c|c|} \hline
	\textbf{Training Data} & \textbf{NISQA} & \textbf{VCC18} & \textbf{IU} & \textbf{PSTN} & \textbf{All}\\ \hline
	NISQA & \cellcolor[HTML]{e6e6e6}0.53* & 1.15 & 3.50 & 0.92 & 2.02 \\ \hline
	VCC18 & 0.98 & \cellcolor[HTML]{e6e6e6} \textbf{0.71*} & 4.19 & 0.99 & 2.36 \\ \hline
	IU & 3.46 & 3.52 & \cellcolor[HTML]{e6e6e6} \textbf{0.48*} & 2.65 & 2.54 \\ \hline
	PSTN & 0.82 & 1.03 & 3.00 & \cellcolor[HTML]{e6e6e6} \textbf{0.51*} & 1.70 \\ \hline
\hline	All & 0.82 & 0.85 & 1.14 & 0.92 & 0.97 \\ \hline
	All (+ BAL) & 0.56* & \underline{0.73*} & 0.63* & 0.53* & \underline{0.59*} \\ \hline
	All (+ MDF) & \underline{0.48*} & 0.76* & 0.73* & \underline{0.52*} & 0.62* \\ \hline
	All (+ MDF + AlignNet) & \textbf{0.47*} & 0.75* & \underline{0.56*} & 0.53* & \textbf{0.57*} \\ \hline
\end{tabular}
    }
    \vspace{-4mm}
\end{table}

\vspace{-2mm}
\subsection{Training regimens and network architectures}
\label{ssec:baselineResults}
\vspace{-1mm}
We demonstrate the improvements provided by MDF and AlignNet through comparison with a series of baselines.
The first set of baselines trained MOSNet on each target dataset individually.
We also trained MOSNet with the conventional approach discussed in Sec.~\ref{conventional}, which is denoted as ``All" in the results tables.
The MDF approach is denoted as ``All (+ MDF)."

We also trained MOSNet using the bias-aware loss (BAL) method defined in~\cite{MittagQomex2021}.
BAL seeks to address the corpus effect in  training by using least-squares to estimate a scale and shift for each dataset after each training epoch. The scale and shift are then used in the loss function to attempt to harmonize the disparate experiment scores.
BAL training relies on a hyperparameter $r_\text{th}$.
Scale and shift are not used in the loss calculation until
training correlation exceeds $r_\text{th}$.
We selected a single threshold of $r_\text{th} = 0.6$ based on curves shown in~\cite{MittagQomex2021}.
We use the same dataset as reference for \alignnet{} and BAL.

We implemented \alignnet{} with MOSNet as the \audionet{} and an \alignmentnet{} that uses a 10-dimensional dataset embedding and has 5 fully connected layers of dimension 16.
MOSNet has roughly 1.2 million parameters and the \alignmentnet{} has 
roughly 1100, 
meaning the \alignmentnet{} was less than 0.1\% of the total network size.
To encourage the \alignmentnet{} to focus only on dataset alignment we froze the pretrained \audionet{} for the first epoch.
In the tables \alignnet{} is denoted as ``All (+ MDF + AlignNet)".
Neither MDF nor AlignNet add measurable additional training time.
BAL increases training time by about 50\%.

\subsection{Results}
\label{ssec:results}
\vspace{-1mm}
As expected, training on individual datasets gives reasonable depth for some of the larger datasets, but almost no breadth. 
Further, only datasets with over 1000 files demonstrate good depth, and some of the particularly small datasets fail to train meaningfully at all.
The conventional approach (``All'' row) offers an improvement over individual training for the smaller datasets --- it gives better breadth as one would expect.
However, outside of VMC22, for datasets with more than 1000 files ``All'' is significantly worse than individual training.
Individual training is particularly successful for FFTNet and NOIZEUS because those datasets contain very unique conditions.

MDF has no specific features to address dataset alignment but it often outperforms or matches the BAL method, including when RMSE and correlations are measured across all the data in the small dataset study (``All'' column in Table~\ref{table:small-corrs}).
The combination of pretraining and a loss function that gives equal weight to each dataset enables this network to reliably estimate scores with no per-dataset side information.

\alignnet{} demonstrates the best performance of all the training regimens and model architectures that are trained on all the datasets, particularly in the small dataset study.
There it achieved the highest LCCs for every dataset but two and the lowest RMSE for every dataset but one. 
It also achieved the best performance in the ``All'' column for both metrics in both studies.
This demonstrates that \alignnet{} is a powerful tool for reconciling different datasets.
Remarkably, when compared to training on individual datasets, \alignnet{} gives better or similar results on \emph{all datasets at once}.
\begin{figure}[h]
    \vspace{-2mm}
    \centering
    \includegraphics[width=\columnwidth]{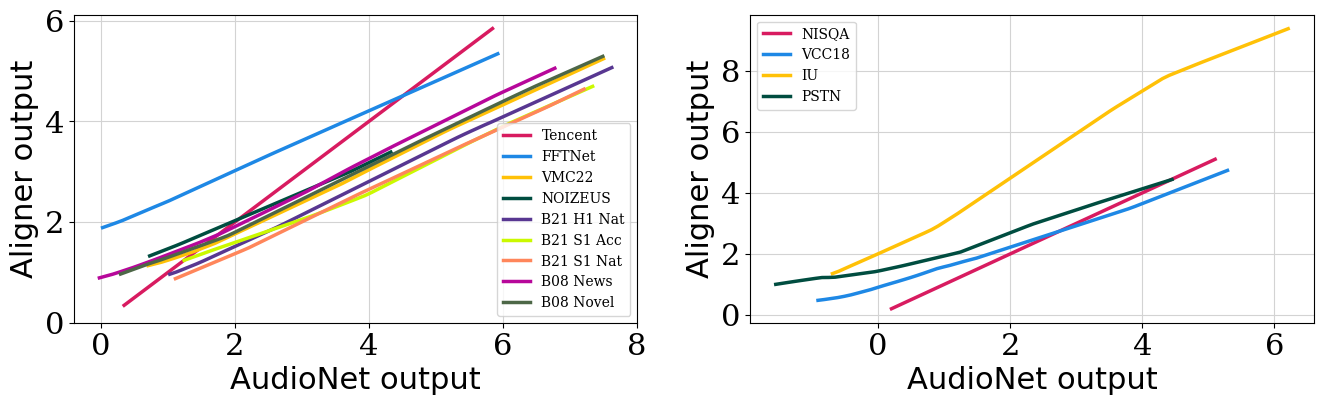}
    \\[-3mm]
    \caption{
    Learned dataset score alignment functions. Left - small dataset study. Right - large dataset study. 
    Functions plotted only over observed values in training data for each dataset.
    }
    \label{fig:learned-alignments}
    \vspace{-7.5mm}
\end{figure}
\alignnet{} is an estimator that is capable of properly ranking a multitude of diverse audio conditions, with an insignificant increase to overall model complexity compared to only using the \audionet{}.
In addition to better estimates, \alignnet{} offers a few other benefits over BAL.
\alignnet{} simultaneously updates the \audionet{} and the \alignmentnet{}, which is far more efficient than the iterative approach of BAL.
Further,~\cite{MittagQomex2021} states that BAL is very sensitive to the $r_\text{th}$ parameter and requires optimization for each dataset, which becomes impractical with a large total number of files.
\alignnet{} has a similar hyperparameter which sets the duration that \audionet{} is frozen once MDF starts, but we consistently see best performance with the fixed value of one epoch.

It is easy to visualize the learned alignments from \alignnet{} by plotting the \alignmentnet{} score estimates vs intermediate score estimates from \audionet{}, as seen in Fig.~\ref{fig:learned-alignments}.
Datasets with similar properties give similar alignment functions, which can be seen clearly in the alignment function plot for the small dataset study.
After they have been learned, these alignments could be approximated by monotonic third-degree polynomials, as previously recommended by~\cite{P1401}.
Finally, note that there can be additional information carried in the range of the intermediate scores of the non-reference datasets.
These scores can extend beyond the nominal range which may speak to the impairments in those datasets relative to those in the reference dataset.

\vspace{-2mm}
\section{Conclusion}
\label{section:conc}
\vspace{-1mm}
\alignnet{} with MDF can reconcile different rated speech attributes such as naturalness, acceptability, and quality.
In the small dataset study four attributes were
successfully harmonized resulting in a more robust NR estimator and revealing the relationships between the different attributes and experiment contexts.
This work demonstrates that disparate scores from distinct listening experiments can be used harmoniously for NR speech estimator training by adding a small alignment network to an existing NR speech estimator.
This type of work is always limited by the data used.  We argue our work is very strong in this regard, but we nonetheless seek to do additional work with even more and broader data. We also plan to experiment with other choices for the \audionet{} used inside of \alignnet{} 
and to further interpret the learned relationships between datasets.
Finally, studying the performance of an \alignnet{} model with MDF on unseen datasets could provide additional insights into the practical use of such a model.

\bibliographystyle{IEEEtran}
\bibliography{sourcesArXiv}

\end{document}